\documentclass[conference]{IEEEtran}
\IEEEoverridecommandlockouts

\usepackage{cite}
\usepackage{amsmath,amssymb,amsfonts, nccmath}
\usepackage{algorithm, algorithmic}
\usepackage{graphicx}
\usepackage{textcomp}
\usepackage{subfig}
\usepackage[]{footmisc}
\usepackage{booktabs}

\newtheorem{definition}{Definition}
\usepackage{xcolor}
\def\BibTeX{{\rm B\kern-.05em{\sc i\kern-.025em b}\kern-.08em
T\kern-.1667em\lower.7ex\hbox{E}\kern-.125emX}}

\newcommand{\Pc}{\mathcal{P}}
\newcommand{\Ec}{\mathcal{E}}
\newcommand{\Sc}{\mathcal{S}}
\newcommand{\Uc}{\mathcal{U}}
\newcommand{\Cc}{\mathcal{C}}
\renewcommand\IEEEkeywordsname{Keywords}

\DeclareMathAlphabet{\pazocal}{OMS}{zplm}{m}{n}

\begin{document}

\title{Bayesian Inference of a Social Graph with \\ Trace Feasibility Guarantees\\

\thanks{This work is funded by the ANR (French National Agency of Research) by the “FairEngine” project under grant ANR-19-CE25-0011.}
}

\author{\IEEEauthorblockN{Effrosyni Papanastasiou}
\IEEEauthorblockA{\textit{Sorbonne University, CNRS, LIP6}\\
F-75005, Paris, France\\
effrosyni.papanastasiou@lip6.fr}
\and
\IEEEauthorblockN{Anastasios Giovanidis}
\IEEEauthorblockA{\textit{Sorbonne University, CNRS, LIP6}\\
F-75005, Paris, France\\
anastasios.giovanidis@lip6.fr}
}

\maketitle

\begin{abstract}

Network inference is the process of deciding what is the true unknown graph underlying a set of interactions between nodes. There is a vast literature on the subject, but most known methods have an important drawback: the inferred graph is not guaranteed to explain every interaction from the input trace. We consider this an important issue since such inferred graph cannot be used as input for applications that require a reliable estimate of the true graph. On the other hand, a graph having trace feasibility guarantees can help us better understand the true (hidden) interactions that may have taken place between nodes of interest. The inference of such graph is the goal of this paper. Firstly, given an activity log from a social network, we introduce a set of constraints that take into consideration all the hidden paths that are possible between the nodes of the trace, given their timestamps of interaction. Then, we develop a non-trivial modification of the Expectation-Maximization algorithm by Newman \cite{b7}, that we call \texttt{Constrained-EM}, which incorporates the constraints and a set of auxiliary variables into the inference process to guide it towards the feasibility of the trace. Experimental results on real-world data from Twitter confirm that \texttt{Constrained-EM} generates a posterior distribution of graphs that explains all the events observed in the trace while presenting the desired properties of a scale-free, small-world graph. Our method also outperforms established methods in terms of feasibility and quality of the inferred graph.

\end{abstract}


\section{Introduction}
\label{intro}

\textit{Network inference}, or \textit{reconstruction} is the problem of predicting the presence or absence of edges between a set of nodes that form the vertices of a graph, given an observed set of data, i.e., the \textit{trace} \cite{b10}. Network inference has long been considered an important task; initially, it attracted a lot of attention in computational biology with various works reconstructing biological networks using representation learning, Bayesian networks, etc \cite{b11}. At the same time, studies applying network inference on social network data began emerging in literature and continue until today, thanks to the rapid growth of Online Social Networks (OSNs) \cite{b1, b2, b3}. The main goal of such works is to infer the \textit{influence} between users - an important property of a social network \cite{b6} - and usually rely on diffusion models that capture the way information is diffused through the network. Such models include the \textit{Independent Cascade (IC)} model, also called \textit{Susceptible-Infected-Recovered (SIR)} in epidemiology \cite{b5, b9} and the \textit{Susceptible-Infected (SI)} model \cite{b4}. The \textit{Linear} or \textit{General Threshold} model has also been extensively used for network inference \cite{b6}. 

After the choice of a diffusion model, the corresponding model parameters can be learned with several approaches, such as maximum likelihood \cite{b6, b8}, Expectation-Maximization (EM) \cite{b5, b7, b9} and other static or continuous-time models \cite{b6}. For example, Saito et al. \cite{b5}, used the IC model and applied the EM algorithm to learn the pairwise transmission probabilities of influence between users. More recently, Bourigault et al. \cite{b9} presented an embedded version of the IC model on OSNs that learns information diffusion probabilities along with the representation of users in the latent space. More recently, Newman \cite{b7} proposed an EM algorithm that is designed for network inference given unreliable data with the help of a set of parameters that estimate the size of the errors. Peixoto \cite{b12} approached network reconstruction similarly to Newman and combined it with community detection, proving that the task of inferring edges can improve the accuracy of community detection and vice-versa.

In the aforementioned inference procedures, we observed that the \textit{feasibility} of the trace in relation to the inferred network is not always guaranteed. This means that the inferred network may not accurately and completely \textit{explain} the input trace. For example, the application of the algorithm proposed by Saito et al. \cite{b5} applied on a Twitter cascade of user tweets and retweets, showed that, while it can predict high diffusion probabilities for some pairs of users and, thus, explain why we observe some retweets in the trace, it cannot do so for every original tweet. As a result, for a considerable number of retweets, we do not know the source of their influence and we, therefore, cannot explain their existence in the trace. The same is true for Newman's network reconstruction algorithm \cite{b7}.

In cases like the above, we say that the trace is not \textit{feasible} with respect to the inferred network. We argue that the concept of trace feasibility is an important condition that an inference framework must meet if we wish to understand the produced networks with relation to the trace and get them to explain every instance of our data. Therefore, we approach the problem of network inference in a novel way by developing a method that infers a posterior distribution of feasible networks that can accurately explain the given trace while respecting the \textit{temporal order} of the observed events (e.g., posts). Towards this goal, we propose a non-trivial modification of the EM inference procedure developed by Newman \cite{b7} by introducing a set of constraints that take into consideration all the (yet unknown) paths that are possible according to the timestamps of interaction between the nodes and therefore derive feasible graphs. 

Our algorithm works for OSN traces that include hundreds of thousands to millions of nodes. We focus the analysis on data from Twitter but we could apply the constraint set on other domains as well. However, to do so, the feasibility constraints should be adjusted each time to the type of data that we work with.

\section{Environment}\label{probdef}

\subsection{Assumptions on the environment}\label{introtrace}

 In the present paper, we will work with traces available from OSNs like Twitter or Weibo. On these platforms, a set of users can generate content that we call \textit{posts}. For example, in the case of Twitter, users can post either \textit{original posts} (i.e., tweets) or \textit{reposts} (i.e., retweets) of original posts from other users. For this paper, we make three crucial assumptions: 
\begin{enumerate}
 \item Users repost from users they follow, i.e., their \textit{followees}. 
 \item The followee from whom each user reposts is included inside the available trace. 
 \item Each user can repost the same post only once.
\end{enumerate}
Of course, the above do not always hold in reality, but they serve as a simplification for our work. As an extension, our method could be modified to include cases where users repost outside the available trace or users, or from users who are not even their followees (e.g., if they traced a tweet from trending topics, hashtags, search, etc). 

For the diffusion of posts, we choose the SI diffusion model from epidemiology \cite{b4}: with respect to each post, a user can transition from the susceptible state (i.e., when one of their followees posts or reposts something) to the infected state (i.e., when the user reposts it) only once and cannot transition back. In addition, we consider that an infected user can influence their yet uninfected followers during all the consecutive timestamps. 

\subsection{Trace description}

Throughout this paper, we use $\Pc$ to denote our \textit{trace} that is a log of T \textit{posts} and \textit{reposts}, generated by a set $\Uc$ of $|\Uc|=$N users on an OSN ($\text{N} \leq \text{T}$). Each line in $\Pc$ is a quadruple $(pid, t,uid,rid)$ that includes four types of information: i) the unique post id $pid$; ii) its timestamp $t$; iii) the $uid \in \Uc$ of the user who posted it; and iv) a repost id $rid$ that is either equal to $-1$ if the post is an original post, or equal to the $pid$ of the original post if it is a repost. All posts in $\Pc$ are ordered according to their timestamps. It is important to underline here that social media logs, like in Twitter do not provide the identity of the user from whom someone found and reposted a post; they only include the $rid$, from which we can then track the original author. 

\section{Problem formulation}\label{preprocess}

\subsection{Problem definition}

Since posts propagate by reposting, we assume the existence of an underlying friendship network between the N different users in the trace that is unknown to us. This network is a directed \textit{social graph} $G = (V, E)$ where the nodes are the users of the trace ($V=\Uc$) and the edges include the friendships between the users. We represent $G$ with an adjacency matrix N$\times$N, denoted by \textbf{A}, where each element $A_{ij}$ is equal to $1$ if user $j$ follows user $i$ and $0$ otherwise. Our intuition is that if a user $j$ shares content frequently from user $i$, it is more probable that $j$ follows $i$. The goal of this work is to infer the unknown friendship network $G$, with trace feasibility guarantees, by inferring the hidden path each original post takes from user to user in the trace. It is important to note here that we can only retrieve friendships between users that have had at least one interaction with each other in the given trace. The richer the trace, the more complete our network inference will be. 

\subsection{Preliminaries}
Our method relies on rich information extracted from a social media trace during the pre-processing phase. We begin by extracting from $\Pc$ the set of original posts denoted by $\Sc$ with cardinality $|\Sc|=$ S. We denote by $r_{s}$ the $uid$ of the user who originally posted each post $s$, i.e., $ (s, t, r_{s}, -1) \in \Pc$ for some $t$.

\begin{definition}[Episode]
For each original post $s \in \Sc$ we define an \textit{episode} as a set of users $\Ec_{s}$ = $r_{s} \cup \{u \in \Uc | \exists (pid,t): (pid, t, u, s) \in \Pc\}$. The whole set of episodes is denoted by $\Ec$ and includes S episodes in total.
\end{definition}

Each episode $\Ec_{s}$ includes the user who originally posted $s$, denoted by $r_{s}$, followed by the users who reposted it, in chronological order. We use $i \stackrel{s}{\prec} j$ to say that user $i$ appears in $\Ec_{s}$ before $j$, and we call this pair a temporally \textit{ordered pair $(i,j)_{s}$}. We count $M_{ij}$ out of the S total episodes where $i \stackrel{s}{\prec} j$. If $M_{ij} > 0$, then it is possible that $j$ has reposted from $i$ an original post or repost and we call this pair an \textit{active pair}. Hence, it is a very important quantity for the inference of the hidden post propagation paths and we will make use of it in the next sections. The total number of active pairs is denoted by L and is equal to $\sum_{i \neq j} 
\mathbf{1}(M_{ij} > 0)$, where $\mathbf{1}(z)=1$ when $z$ is true. All information extracted from the trace is summarized in Table \ref{tab1}. 

\begin{table}[htbp]
\captionsetup{justification=centering, labelsep=newline}
\caption{\textsc{Trace Information}}
\begin{center}
\begin{tabular}{|c|c|}
\hline
\textbf{Symbol}&{\textbf{Definition}} \\
\hline
$\Pc$ & Set of posts and reposts in the trace, $|\Pc|=$ T\\ \hline
$\Sc$ & Set of original posts in the trace, $|\Sc|=$ S\\ \hline
$\Uc$ & Set of users that posted or reposted a post \\ \hline
N & Number of different users appearing in the trace \\ \hline
$\Ec$ & Set of episodes, $|\Ec|=$S \\ \hline
$\Ec_{s} \in \Ec$ & Episode of post $s$, $1 \leq s \leq$ S \\ \hline
$r_{s}$ & The id of the user that originally posted $s$ \\ \hline
$ i \stackrel{s}{\prec} j$ & User $i$ reposted or posted $s$ before $j$ \\ \hline
$M_{ij}$ & Number of episodes where $i \stackrel{s}{\prec} j$ \\ \hline
L & Number of active pairs with M$_{ij} > 0$\\
\hline
\end{tabular}
\label{tab1}
\end{center}
\end{table}

\subsection{Diffusion model per episode}

Given an episode $\Ec_{s}$ and an ordered pair $(i,j)_{s}$, we first define the value $X_{ij}(s) \in \{0,1\}$ that is equal to $1$ if user $j$ reposted $s$ directly from $i$ (i.e., post $s$ propagated from $i$ to $j$) and equal to $0$ otherwise. However, the real value of $X_{ij}(s)$ is still unknown to us. Therefore, for the given post $s$, if we look into the temporal order of its reposts, we could think of several feasible ways or \textit{paths} through which the post could have spread to reach the users that reposted it. These paths form a \textit{propagation graph} $G_{s}=\{V_{s}, E_{s}\}$ per episode, with the users in episode $\Ec_{s}$ as nodes ($V_{s}=\Ec_{s}$), and the edges set $E_{s}$ containing the edges that are activated for the given post. Each activated edge follows the direction of
propagation, e.g., an edge $(i,j)$ in $G_{s}$ means that $X_{ij}(s)=1$. 

As mentioned before, it is crucial that the paths in $G_{s}$ take into account the temporal order of the users' posts and reposts. For example, we can think of a hypothetical episode $\Ec_{s}$ of post $s$, depicted in Fig. \ref{fig2}. The user \textit{Frosso (Fr)} was the first who posted $s$, followed by \textit{Phoebe (Ph)} and then, \textit{Anastasios (An)}. Therefore, the possible propagation graphs, based on the chronological ordering of each repost in the episode, are the following:
\begin{itemize}
\item Users \textit{Phoebe} and \textit{Anastasios} reposted it both directly from \textit{Frosso}. This corresponds to a propagation graph $G_{s}$ with two directed paths: i) from the node \textit{Frosso} to the node \textit{Phoebe}; and ii) from the node \textit{Frosso} to the node \textit{Anastasios} (case I in Fig. \ref{fig2}).
\item \textit{Phoebe} reposted it from \textit{Frosso} and then \textit{Anastasios} from \textit{Phoebe}. This corresponds to a propagation graph $G_{s}$ with one directed path: from \textit{Frosso} to \textit{Phoebe} and then to \textit{Anastasios} (case II in Fig. \ref{fig2}).
\end{itemize} 

Notice that each path in each possible tree follows the time-ordering of reposts. In addition, $G_{s}$ is an \textit{arborescence} with root $r_{s}$ \cite{b13}: this means that (i) $G_{s}$ is a DAG; and (ii) given the vertex $r_{s}$ called the root and any other vertex $u$ in $G_{s}$, there is exactly one directed path from $r_{s}$ to $u$. Equivalently, $G_{s}$ is a directed, rooted tree in which all edges point away from the root.

\section{Feasibility and constraints}\label{constrs}

\subsection{Feasibility definition}
\label{feas}
Given our problem definition, for each original post in $\Sc$, we need to infer a propagation DAG that is \textit{feasible} in relation to the trace. 

\begin{figure}[htbp]
\centerline{\includegraphics[width=0.3\textwidth]{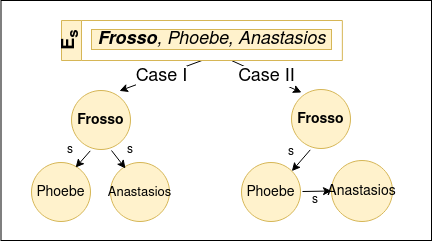}}
\caption{Possible propagation graphs for episode $\Ec_{s}$.}
\label{fig2}
\end{figure}

\begin{definition}[Feasible propagation DAG per episode] \label{def2}
Given an episode $\Ec_{s}$ from the trace, we say that a propagation DAG $G_{s}=\{V_{s}, E_{s}\}$ is \textit{feasible in relation to $\Ec_{s}$}, if $V_{s}=\Ec_{s}$ and there exists (at least) one directed path from the root user $r_{s}$ to every other user $j \in V_{s} \backslash r_{s}$. For each edge $(i,j)$ of the path it holds that $i \stackrel{s}{\prec} j$ in the trace. 
\end{definition}

Using a DAG for each episode, $G_{s}$ with $s=1,2,.., \text{S}$, we can construct the full adjacency matrix \textbf{A} of the final friendship graph $G$ as follows: we set $A_{ij}=1$ if there exists at least one propagation DAG $G_{s}$ where the edge $(i,j)$ exists, and $0$ otherwise.

\begin{definition}[Feasible friendship graph]
We define the adjacency matrix \textbf{A} of an inferred network $G$ as \textit{feasible} in relation to an OSN trace if, for every original post $s$ there exists a subgraph in $G$, which is a feasible propagation graph $G_{s}$ as defined in Def. \ref{def2}.
\end{definition} 

\begin{figure}[htbp]
{\includegraphics[width=0.57\textwidth]{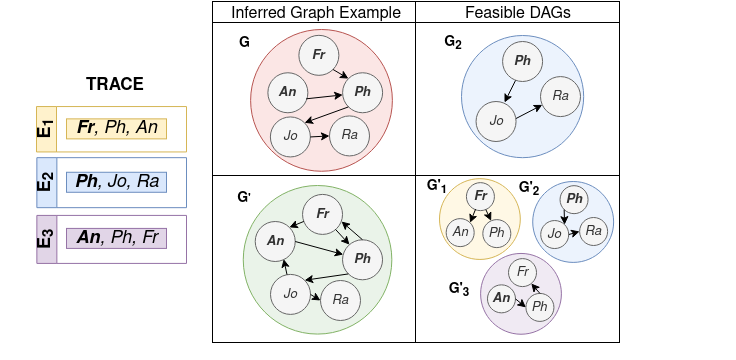}}
\caption{Example of feasibility check given a trace.}
\label{fig3}
\end{figure}

For example, in Fig. \ref{fig3}, given a trace of three episodes $\Ec=\{\Ec_{1},\Ec_{2},\Ec_{3}\}$ that involves five users \textit{Frosso (Fr), Anastasios (An), Phoebe (Ph), Rachel (Ra), Joey (Jo)}, we can see the case of a non-feasible $G$ and a feasible $G'$. $G$ is non-feasible since there exists no feasible path from source node \textit{Frosso} to \textit{Anastasios} for the case of episode $\Ec_{1}$ and there exists no feasible path from source node \textit{Anastasios} to \textit{Frosso} for the case of episode $\Ec_{3}$. In contrast, $G'$ is feasible since there exists a feasible propagation graph for each episode. 

\subsection{Feasibility constraints on retweets behavior}
The real value of $X_{ij}(s)$ defined in Section \ref{preprocess} is not available, but we can limit the possible combinations by introducing a set of constraints on $X_{ij}(s)$, that ensure that a user $j \in \Ec_{s}$ has reposted $s$ directly by at least one user $i \in \Ec_{s}$ who has reposted $s$ earlier according to the trace. As a result, for each episode $\Ec_{s} \in \Ec$, and each user $j \in \Ec_{s} \backslash \{r_{s}\}$, the constraints take the following form: 
\begin{equation}
\sum_{i \in \Ec_{s} \text{ s.t. } i \stackrel{s}{\prec} j}{X}_{ij}(s) \geq 1, \forall j \in \Ec_{s} \backslash \{r_{s}\}.
\label{eq1}
\end{equation}
For all episodes we will get T-S constraints in total, i.e., as many constraints as the number of reposts with $\sum_{s=1}^{\text{S}} (|\Ec_{s}| - 1)\cdot(|\Ec_{s}|)/2$ unknown variables in total. The way we formed the constraints, we allow the possibility that a user $j$ has reposted the post $s$ from more than one users instead of only one user for reasons that will become clear in the following section. For example, for episode $\Ec_{s}$ in Fig. \ref{fig2} we will get constraints: $X_{FrossoPhoebe}(s) \geq 1$ and $X_{FrossoAnastasios}(s)$ $+$ $X_{PhoebeAnastasios}(s) \geq 1$. This means that the user $Phoebe$ has definitely reposted $s$ from $Frosso$ and that $Anastasios$ has reposted it from $Frosso$, $Phoebe$, or both.

\subsection{Diffusion probabilities}

To be able to infer $X_{ij}(s)$ we make an important assumption: For every ordered pair $(i,j)_{s}$ in an episode $\Ec_{s}$, user $j$ reposted post $s$ from $i$ independently of other episodes with unknown probability $\sigma_{ij} \in [0,1]$ that is common for all episodes. In other words, $X_{ij}(s)$ is an independent Bernoulli random variable with mean $\sigma_{ij}$ that is independent of $s$. 

Our choice to model the uncertainty about the path by using an independent Bernoulli random variable with a fixed mean $\sigma_{ij}$, assumes that user $j$ does not have a contextual behavior that depends on the content of the episode, but rather behaves with randomness when choosing their sources of information. This serves as a simplification for the inference; as an extension, it could be further refined to include context-dependent mean values for different types of episodes. The value of $\sigma_{ij}$ can be seen as the limiting frequency that user $j$ reposts directly a post from $i$ when the number of episodes goes to infinity. Given an ordered pair $(i,j)_{s}$, $\sigma_{ij}$ is equal to: 
\begin{equation}
\sigma_{ij} = \mathbb{E}\left[X_{ij}(s)\right].
\label{eq3}
\end{equation}

Given our intuition that the number of times a user $j$ reposts a user $i$ is indicative of their friendship we introduce to the problem the quantity $Y_{ij}$, as the (unknown) number of times that \textit{j} reposted from $i$, out of the $M_{ij}$ possible ones (the number of times they appear as an active pair). That is:
\begin{equation}
Y_{ij} = \sum_{m=1}^{M_{ij}}X_{ij}(s).
\label{eq2}
\end{equation} 
Since $Y_{ij}$ is the sum of $M_{ij}$ independent Bernoulli random variables with mean value $\sigma_{ij}$, $Y_{ij}$ is an independent Binomial random variable with mean value $M_{ij}\sigma_{ij}$. That is: 
\begin{equation}
 \mathbb{E}[Y_{ij}] = \sum_{m=1}^{M_{ij}}\mathbb{E}[X_{ij}(s)] = \sum_{m=1}^{M_{ij}}\sigma_{ij} = M_{ij}\sigma_{ij}. 
 \label{eq7d}
\end{equation} 
The value of $\sigma_{ij}$ plays an important role in the inference of the relationship between $i$ and $j$, as we will demonstrate in the following sections.

\subsubsection{Constraints on diffusion probabilities $\sigma_{ij}$} 
\label{constr}
Having transformed the problem from solving over $X_{ij}(s)$ to solving over $\sigma_{ij}$, we can constrain $\sigma_{ij} = \mathbb{E}\left[X_{ij}(s)\right]$ to be inside a specific set of values. If we take the expectation of the constraints in \eqref{eq1}, for each episode $\Ec_{s} \in \Ec$, and each user $j \in \Ec_{s} \backslash \{r_{s}\}$, we end up with the following set of constraints on parameters $\sigma_{ij}$:
\begin{align}
& \sum_{i \in \Ec_{s} \text{ s.t. } i \stackrel{s}{\prec} j}\sigma_{ij} \geq 1, \forall j \in \Ec_{s} \backslash \{r_{s}\} \label{eq4a}\\
& \sigma_{ij} \in [0,1], \text{ } \forall (i,j) \in \Uc.
\label{eq4}
\end{align}
We define with F$_{\boldsymbol{\sigma}}$ the feasibility space of the parameters vector $\boldsymbol{\sigma}$ that includes all $\sigma_{ij} \text{ parameters}, (i,j)\in \Uc, \text{ such that (5) and (6) hold.}$ In this case, for episode $\Ec_{s}$ in Fig. \ref{fig2} the constraints change to the following: $\sigma_{FrossoPhoebe} \geq 1$ and $\sigma_{FrossoAnastasios}$ $+$ $\sigma_{PhoebeAnastasios} \geq 1$. This means that user $Phoebe$ has reposted $s$ from $Frosso$ with probability $\sigma_{FrossoPhoebe} = 1$ and that the probabilities of $Anastasios$ reposting it from $Frosso$ and $Phoebe$ must sum up to a value inside the interval $\in [1,2]$. 

As a result, the parameters $\sigma_{ij} \in [0,1]$ are the problem unknowns that replace the $X_{ij}(s) \in \{0,1\}$ for all episodes where an action from user $i$ precedes an action from $j$. For the whole trace $\Ec$, we will get a set of constraints $\Cc = \{c_{1}, c_{2}, ..., c_{(\text{T-S})}\}$, where, each element $c_{k} \in \Cc, 1 \leq k \leq$ (T-S), corresponds to the constraint of a $(r_{s},j)$ tuple, where $r_{s} \in \Ec_{s}$ and user $j \in \Ec_{s} \backslash \{r_{s}\}$, and is defined by \eqref{eq4a}-\eqref{eq4}. By imposing this set of constraints on the parameters $\sigma_{ij}$, we have drastically reduced the number of our problem's unknowns to the number of possible $(i,j)$ pairs from the users set $\Uc$, i.e., we now have $\text{N}(\text{N}-1)$ unknowns. 

\subsubsection{Removing redundant constraints}
\label{rem-con}
We notice that given a trace, some constraints become redundant and can be removed according to the following rules: 
\begin{itemize}
 \item If all parameters $\sigma_{ij}$ that are included in a constraint $c_{k} \in \Cc$, are also included in a different constraint $c_{w} \in \Cc$, then $c_{w}$ is removed from $\Cc$. 
 \item In \eqref{eq4a}, we observe that the first constraint of each episode includes only one variable, which is the $\sigma_{ij}$ between the first user $i=r_{s}$ and the second user $j$ in the episode. Therefore, given also that $\sigma_{ij} \in [0,1]$, all parameters between the first and the second user of each episode become $\sigma_{ij}=1$. As a result, the first constraint per episode is removed, since the solution for these parameters has already been found. 
\end{itemize}
Note that for $M_{ij}=0$, $\sigma_{ij}=0$. Generally, the exact number of constraints by which our problem will be reduced depends on the characteristics of each trace.

\section{Problem Modeling and Learning Method}\label{modeling}

As mentioned in the introduction, we develop a non-trivial modification of Newman's EM algorithm proposed in \cite{b7} that was designed for network inference given erroneous data. In our case, our data is not erroneous, but rather incomplete; however, we take advantage of Newman's probabilistic modeling and we adapt its parameters and the EM equations to our case. 

\subsection{Parameters}

Firstly, in a similar fashion to Newman \cite{b7}, we assume that the relationship between the underlying network $G$ and the trace can be expressed in the form of a probability function $\textit{P}(\text{data}|\textbf{A},\theta)$, which is the probability of generating the particular trace $\Pc$, given the adjacency matrix \textbf{A} and a set of additional model parameters, denoted by $\theta$. The parameters $\theta$, added to cover a larger range of possibilities for the type of graph and the way the data is generated,  are the following: 
\begin{enumerate}
\item To model our uncertainty about the structure of the graph $G$, we assume a uniform prior probability $\rho$
of the existence of an edge in any position between any pair of nodes, i.e. $G$ has been drawn under the Erdős–Rényi model with parameter $\rho$.
\item The values $\sigma_{ij}$, which is the fixed probability that $j$ shares content from $i$.
\item The \textit{true-positive utilization rate $\alpha$}: the probability of post propagation
through existing edges of the underlying network $G$.
\item The \textit{false-positive utilization rate $\beta$}: the probability of post propagation through non-existing edges of the underlying network $G$.
\end{enumerate}
We see that $\alpha$ and $\beta$ are global parameters, conditioned on the existence or not of an edge in the ground truth network $G$.

\subsection{Learning Method}\label{learning}
To find the most probable value of the parameters $\theta$ given the observed data and infer a graph with maximum likelihood, we will develop an application of Expectation-Maximization (EM): an iterative algorithm designed to find the maximum a posteriori (MAP) estimates of parameters in statistical models that depend on unobserved latent variables. Each EM iteration will alternate between two steps: i) an expectation (E) step, which creates the expectation of the log-likelihood using the current estimate for the parameters $\theta$; and ii) a maximization (M) step, which finds the parameters that maximize the expected log-likelihood of the E-step. The estimated parameters are then used in the next E-step and so on until convergence is reached. 

We begin in the same way as Newman \cite{b7} and we apply the Bayes' rule: 
\begin{equation}
\textit{P}(\textbf{A},\theta|\text{data}) = \frac{\textit{P}(\text{data}|\textbf{A}, \theta)\textit{P}(\textbf{A}|\theta)\textit{P}(\theta)}{\textit{P}(\text{data})}.
\label{eq5}
\end{equation}

The probability that we get the specific set of reposts, given \textbf{A} and the parameters $\theta=$\{$\alpha, \beta, \rho, \boldsymbol{\sigma}$\}, differs here from Newman since we have introduced the hidden number of interactions between users, $Y_{ij}$. Given the ordered nodes of an episode, each repost path is chosen independently per episode. In addition, we assumed as prior knowledge that between any two
nodes in \textbf{A} an edge has been drawn with probability $\rho$. Therefore we get:

\begin{align}
& \textit{P}(\text{data}|\textbf{A}, \theta)\textit{P}(\textbf{A}|\theta) = \prod_{i \neq j}{\left[ \alpha^{Y_{ij}}{(1 - \alpha)}^{M_{ij} - Y_{ij}}\rho\right]}^{A_{ij}} \ \nonumber 
\\ 
& \times {\left[\beta^{Y_{ij}}{(1 - \beta)}^{M_{ij} - Y_{ij}}(1-\rho)\right]}^{1-A_{ij}}.
\label{eq8}
\end{align}

Given this type of modeling, when $A_{ij}=1$, the $Y_{ij}$ out of the $M_{ij}$ experiments are successful, each with probability $\alpha$. When $A_{ij}=0$, the $Y_{ij}$ out of $M_{ij}$ experiments are successful, each with probability $\beta$. For the whole set of parameters $\theta$, we assume a uniform prior probability $\textit{P}(\theta)$. If we sum \eqref{eq5} over all possible networks \textbf{A}, we find that $\textit{P}(\theta|\text{data}) = \sum_{\textbf{A}} \textit{P}(\textbf{A}, \theta| \text{data})$. Then, as suggested by Newman \cite{b7}, we can apply the well-known Jensen's inequality on the $\log$ of $\textit{P}(\theta|\text{data})$:
\begin{equation}
\log \textit{P}(\theta|\text{data}) = \log \sum_{\textbf{A}} \textit{P}(\textbf{A}, \theta|\text{data}) \geq \sum_{\textbf{A}} q(\textbf{A}) \log \frac{\textit{P}(\textbf{A}, \theta|\text{data})}{q(\textbf{A})}
\label{eq7a}
\end{equation}
where $q(\textbf{A})$ is any probability distribution over networks \textbf{A} satisfying $\sum_{\textbf{A}}q(\textbf{A}) = 1$. We also define the posterior probability of an edge existing between $i$ and $j$ by $Q_{ij} = \textit{P}(A_{ij} = 1| \text{data}, \theta) = \sum_{\textbf{A}}q(\textbf{A})A_{ij}$.

For the E-step, we modify the Newman algorithm by taking the expectation over the set of random variables $Y_{ij}$ at both sides of \eqref{eq7a}: 
\begin{align}
 & \mathbb{E}[\log\textit{P}(\theta|\text{data})]
 \geq \mathbb{E} [\sum_{\textbf{A}} q(\textbf{A}) \log \frac{\textit{P}(\textbf{A}, \theta|\text{data})}{q(\textbf{A})}] \nonumber \\
 & = \sum_{\textbf{A}} q(\textbf{A})\big(\mathbb{E}[\log \textit{P}(\textbf{A}, \theta|\text{data})] - \log q(\textbf{A}) \big).
\label{eq7b}
\end{align}

To find $\mathbb{E}[\log \textit{P}(\textbf{A},\theta|\text{data})]$, we replace \eqref{eq8} into \eqref{eq5}. Setting $\Gamma=\textit{P}(\theta)/\textit{P}(\text{data})$, the expectation of the log of \eqref{eq5} becomes:
\begin{align}
 & \mathbb{E}[\log \textit{P}(\textbf{A},\theta|\text{data})] = 
 log \Gamma + \sum_{i \neq j} \Big[{A_{ij}} \Big(\log\rho + {\mathbb{E}[Y_{ij}]}\log\alpha + \nonumber \\
 & +(M_{ij}-\mathbb{E}[Y_{ij}])\log{(1 - \alpha)}\Big) 
 \nonumber +(1-A_{ij})\Big(\log(1-\rho) + \nonumber\\
 & + {\mathbb{E}[Y_{ij}]}\log\beta +(M_{ij} - \mathbb{E}[Y_{ij}])\log{(1 - \beta)\Big)\Big]}.
\label{eq7c}
\end{align}
Then, by replacing \eqref{eq7d} into \eqref{eq7c}, and then \eqref{eq7c} into \eqref{eq7b}, we get:
\begin{align} \label{eq7f}
\mathbb{E}[\log\textit{P}(\theta|\text{data})] \geq \sum_{\textbf{A}} q(\textbf{A})\log\frac{D_{ij}}{q(\textbf{A})}
\end{align}

\begin{multline} \label{eq7fb}
\text{where, } D_{ij} = \Gamma \prod_{i \neq j}{\left[ \rho \alpha^{M_{ij}\sigma_{ij}}{(1 - \alpha)}^{M_{ij}(1-\sigma_{ij})}\right]}^{A_{ij}} \\
\times {\left[(1-\rho)\beta^{M_{ij}\sigma_{ij}}{(1 - \beta)}^{M_{ij}(1-\sigma_{ij})}\right]}^{1-A_{ij} }.
\end{multline}

For the M-step of the EM algorithm, the function that we want to maximize is $\mathbb{E}[\log\textit{P}(\theta|\text{data})]$. To do so, we need to find the unknown values, $q(\textbf{A})$ and $\theta=$\{$\alpha, \beta, \rho, \boldsymbol{\sigma}$\}, that maximize the expectation on the left-hand side of \eqref{eq7f}, under the feasibility constraints on the parameters set $\theta$. From these, only the $\sigma_{ij}$ have an important constraint set, specified in \eqref{eq4a} and \eqref{eq4}.

\subsection{Solution}
\subsubsection{With respect to $q(\textbf{A})$} 

We notice that the choice of $q(\textbf{A})$ that achieves equality (i.e. maximizes the right-hand side) in \eqref{eq7f} is: 
\begin{equation}
q(\textbf{A}) = \dfrac{D_{ij}}{\sum_{\textbf{A}}D_{ij}}.
\label{eq9}
\end{equation} 
From \eqref{eq9}, in a similar fashion to Newman's method [Eq. (13), 20], and because $\Gamma$ cancels out, we get:
\begin{equation}
q(\textbf{A}) = \prod_{i \neq j}Q_{ij}^{A_{ij}}(1-Q_{ij})^{1-A_{ij}}
\label{eq22}
\end{equation}
where $Q_{ij}$ is the posterior probability that there exists an edge between $i$ and $j$: 
\begin{equation}
 \medmath{Q_{ij} = \dfrac{\rho \alpha^{M_{ij}\sigma_{ij}} (1-\alpha)^{M_{ij}(1-\sigma_{ij})}}{\rho \alpha^{M_{ij}\sigma_{ij}} (1-\alpha)^{M_{ij}(1-\sigma_{ij})} + (1-\rho) \beta^{M_{ij}\sigma_{ij}} (1-\beta)^{M_{ij}(1-\sigma_{ij})}}}.
 \label{eq23}
\end{equation} 
The expression here is also different from Newman, since in the exponents we get the expected number of events (using $M_{ij}\sigma_{ij}$) instead of the number of times $j$ reposts from origin $i$ directly (that is provided directly by the data). Notice also that for $M_{ij}=0$, $Q_{ij}$ becomes equal to the prior probability $\rho$. Moreover, from \eqref{eq9} we observe that $q(\textbf{A})$ is the posterior probability distribution over all possible networks \textbf{A}, $\textit{P}(\textbf{A},\theta|\text{data})$ when $Y_{ij}$ is replaced by its expected value $M_{ij}\sigma_{ij}$. 

\subsubsection{With respect to $\sigma_{ij}$} 
Our goal now is to find the parameters $\theta$ that maximize the right-hand size of \eqref{eq7f}, given the maximising distribution for $q(\textbf{A})$ in \eqref{eq9}, hence given the values of $Q_{ij}$ in \eqref{eq22}. If we take into account that $Q_{ij} = \sum_{\textbf{A}}q(\textbf{A})A_{ij}$ and also that $\sum_{\textbf{A}} q(\textbf{A}) = 1 $, by rearranging the right-hand side of \eqref{eq7f}, the problem becomes equivalent to maximizing:

\begin{align}
& \sum_{\textbf{A}} q(\textbf{A}) \sum_{i \neq j} \sigma_{ij} M_{ij}\left(A_{ij}\log\frac{\alpha}{1-\alpha} + (1-A_{ij})\log\frac{\beta}{1-\beta}\right) \ \nonumber 
\\ 
& = \sum_{i \neq j} \sigma_{ij} M_{ij}\left(Q_{ij}\log\frac{\alpha}{1-\alpha}+ (1-Q_{ij})\log\frac{\beta}{1-\beta}\right).
\label{eq16}
\end{align}
Finally, if $\boldsymbol{\sigma}$ is a vector of size L that includes all $\sigma_{ij}$ instances, we end up with the following linear optimization problem:
\begin{align}
& \max_{\sigma} \sum_{i \neq j}\sigma_{ij}(W_{ij} - \lambda) \label{eq17a}\\
& \text{s.t. } \boldsymbol{\sigma} \in \textit{F}_{\boldsymbol{\sigma}} \nonumber \\
& \text{where } W_{ij} = M_{ij}\left(Q_{ij}\log\frac{\alpha}{1-\alpha} + (1-Q_{ij})\log\frac{\beta}{1-\beta}\right) \nonumber \\
& \text{and } \lambda > 0 \text{ some given penalty for regularisation.}
\label{eq17}
\end{align} 
Our goal is to infer a graph that is feasible and also has the minimum possible number of edges; this is why we added the value $\lambda$ as a penalty into the maximization goal per each iteration. Without it, all $(i,j)$ pairs with $W_{ij} > 0$ would immediately get their $\sigma_{ij}=1$, leading to the inference of more edges than necessary. Therefore, we choose to set $\lambda$ equal to the largest $W_{ij}$ value, i.e. $\lambda = \max_{(i,j) \in W} W_{ij}$. This choice of $\lambda$ forces the optimization goal to be negative and thus, to be guided only by the provided constraints. It is equivalent to penalizing the total expected number of inferred edges.

\subsubsection{With respect to $\alpha, \beta, \rho$}

Next, we maximize the right-hand side of \eqref{eq7f} in terms of parameter $\alpha$ by differentiating it with respect to $\alpha$ and then setting it equal to zero (while holding $\sigma_{ij}$, $q$ constant):
\begin{equation}
 \sum_{i \neq j} Q_{ij} M_{ij} \left(\frac{\sigma_{ij}}{\alpha} - \frac{1-\sigma_{ij}}{1-\alpha}\right) = 0.
 \label{eq18}
\end{equation}
After rearranging, we get:
\begin{equation}
 \alpha = \dfrac{\sum_{i\neq j} M_{ij} \sigma_{ij}Q_{ij}}{\sum_{i\neq j} M_{ij} Q_{ij}}.
 \label{eq19}
\end{equation}
Similarly for $\beta$ and $\rho$, we get:
\begin{align}
 & \beta = \dfrac{\sum_{i\neq j}M_{ij}\sigma_{ij}(1-Q_{ij})}{\sum_{i\neq j}M_{ij}(1-Q_{ij})}, \label{eq20} \\
 & \rho = \dfrac{1}{\text{N(N}-1)} {\sum_{i\neq j} Q_{ij}}
 \label{eq21}
\end{align}
where N is the number of total different users in the trace. 

Finally, we end up with an iterative EM algorithm that iterates between finding an optimal value for q, i.e. a value that allows for \eqref{eq7f} to hold with equality (E-step), and then holding it constant to maximize the right-hand side of \eqref{eq7f} (and therefore also the expectation in the left-hand side of \eqref{eq7f}) with respect to $\theta$ (M-step), through the updates in \eqref{eq17a}, \eqref{eq17}, \eqref{eq19}, \eqref{eq20}, \eqref{eq21}. Our algorithm converges when the L2 norm of improvement $||\textbf{Q}_{new} - \textbf{Q}_{old} || < \epsilon $ falls under some threshold $\epsilon$ that we choose in advance, where \textbf{Q} is the matrix containing all the $Q_{ij}$ values.

\section{Experimental Evaluation}
\subsection{Dataset}
To evaluate our approach we use a real-world Twitter dataset coming from Kaggle, referred to as \textit{Russian} \footnote{https://www.kaggle.com/borisch/russian-election-2018-twitter}. It contains almost 2 million tweets and retweets emitted from 181,621 users during the Russian presidential elections of 2018. Users are anonymous and tweets are ordered in time. We choose the first 500,000 lines of the trace. Each line is a quadruple \textit{[PostID, TimeStamp, UserID, RePostID]}. We remove all the tweets that have not been retweeted by any users and all the retweets for which we do not know the user who originally posted it. In addition, we delete retweets that appear more than once for the same user and tweet. The final statistics are summarised in Table \ref{tab3}. 

\subsection{Experimental Settings}

\subsubsection{Environment} We run the experiments on a Google Cloud virtual instance with 16 vCPUs and 128 GB RAM. For the solution of the optimization problem, we use PuLP \footnote{https://pypi.org/project/PuLP/}, an open-source linear programming library for Python.

\subsubsection{Number of constraints} 
For a trace set of size $|\Pc|=216,989$ the number of constraints is $198,543$. After removing the redundant ones according to Section \ref{rem-con}, we observe an approximately 10\% percent decrease in the number of constraints ($=170,209$). However, the level of decrease generally depends on the sparsity of the trace network itself. The more connected the initial network, the higher the decrease we observe.  

\begin{table}[]
\captionsetup{justification=centering, labelsep=newline}
\label{tab-constr}
\centering
\caption{\textsc{Basic statistics on Russian after pre-processing}}
\vspace{-2mm}
\begin{tabular}{l|l|}
 & \textbf{Russian} \\ \hline
\multicolumn{1}{|l|}{\textbf{Time window}} & 20 days \\ \hline
\multicolumn{1}{|l|}{\textbf{Trace size $|\Pc|$}} & 216,989 \\ \hline
\multicolumn{1}{|l|}{\textbf{\#original tweets}} & 14,781 \\ \hline
\multicolumn{1}{|l|}{\textbf{\#retweets}} & 202,208 \\ \hline
\multicolumn{1}{|l|}{\textbf{\#users}} & 42,011 \\ \hline
\multicolumn{1}{|l|}{\textbf{\% users with \#tweets \textgreater 0}} &  13.00 \\ \hline
\multicolumn{1}{|l|}{\textbf{\% users with \#retweets \textgreater 0}} &  94.30 \\ \hline
\end{tabular}
\label{tab3}
\vspace{-2mm}
\end{table}

\subsubsection{Initialization and convergence rule}

Parameters $\alpha$, $\beta$ and $r$ are initialized randomly. The threshold $\epsilon$ of our algorithm's convergence criterion on the L2 norm $||\textbf{Q}_{new} - \textbf{Q}_{old} || < \epsilon$ is set equal to $\epsilon = 0.001$.

\begin{figure}%
  \subfloat{{\includegraphics[width=4.5cm]{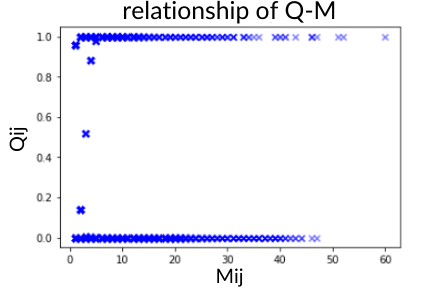} }}%
  \subfloat{{\includegraphics[width=4.5cm]{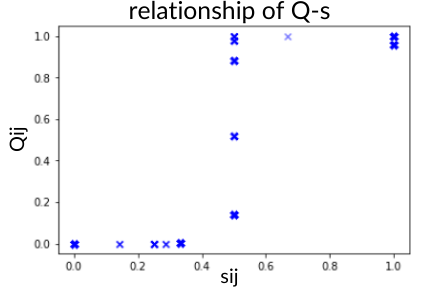} }}%
  \caption{\texttt{Constrained-EM} results.}%
  \label{fig:results}%
\end{figure}

\section{Results and Comparison}

Our algorithm \texttt{Constrained-EM} takes 109 iterations and approximately 60 hours to converge. The converged parameters of our method, are
$\alpha^{*}= 0.9932, \beta^{*}= 0.0001, r^{*}= 0.0034$. This means that there is an approximately $99\%$ probability that a post propagated through an edge present in the inferred network $G$. The small value of $\beta$ suggests that there are very few
false-positive utilized edges: a post from the trace propagates through an edge where none exists around $0.001\%$ of the time.

In Fig. \ref{fig:results} (left), we show the final values of $Q_{ij}$ in relation to the number of times $M_{ij}$ that each $(i,j)$ edge is observed in the trace. As we can see, when $M_{ij}$ is relatively small, $Q_{ij}$ alternates between the whole range of $[0,1]$. As $M_{ij}$ becomes larger, the $Q_{ij}$ is either $0$ or $1$ and finally, for the larger values of $M_{ij}$, the value of $Q_{ij}$ stabilizes to $1$. This could be attributed to the fact that the more times an edge is observed in a trace, the more certain we become about the existence of the edge; equivalently the more times a user retweets after some other user, the more certain we become about the existence of a friendship between them. Moreover, the different $Q_{ij}$ results for the same $M_{ij}$ values depict the important role the constraints play in the inference process. 
Regarding the $\sigma_{ij}$ values, we observe in Fig. \ref{fig:results} (right) that for $\sigma_{ij} > 0.5$, $Q_{ij}$ is almost $1$ and for $\sigma_{ij} < 0.5$, $Q_{ij}$ becomes $0$. For $\sigma_{ij} = 0.5$, $Q_{ij}$ alternates between values in [$0.1, 1$]. Hence, we confirm our intuition that the probability that user $j$ follows $i$ depends on the probability $\sigma_{ij}$ that $j$ reposts $i$.

\begin{figure*}[htbp]
  \subfloat[\centering \texttt{Constrained-EM}]{{\includegraphics[width=3.5cm]{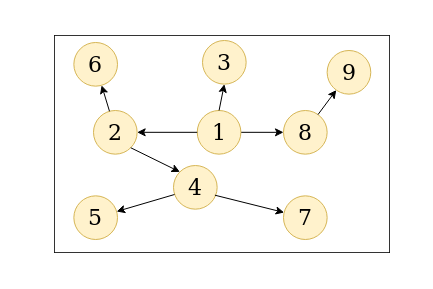} }}%
  \subfloat[\centering \texttt{Saito et al.} \cite{b5}]{{\includegraphics[width=3.5cm]{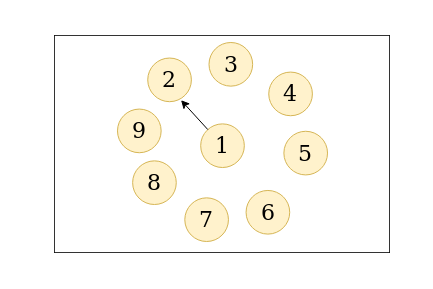} }}%
  \subfloat[\centering \texttt{Star}]{{\includegraphics[width=3.5cm]{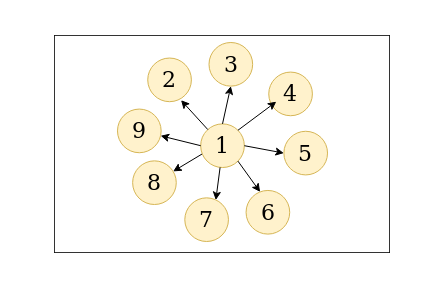} }}%
  \subfloat[\centering \texttt{Chain}]{{\includegraphics[width=3.5cm]{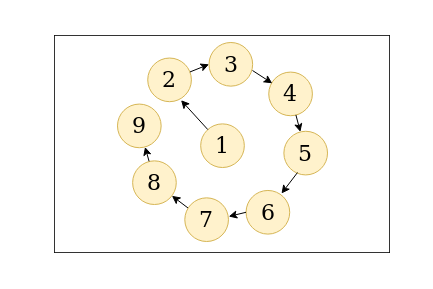} }}%
  \subfloat[\centering \texttt{Newman} \cite{b7}]{{\includegraphics[width=3.5cm]{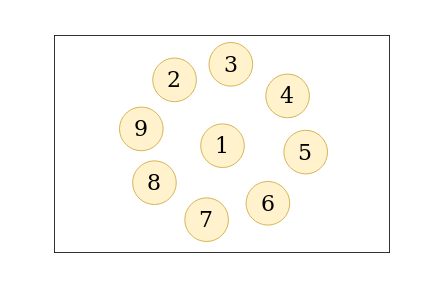} }}%
  \caption{Propagation graph inference for an episode $\Ec_{s}=\{1,2,...,9\}$ from the trace.}%
  \label{fig:example}%
\end{figure*}

\begin{table*}[htbp]
\captionsetup{justification=centering, labelsep=newline}
\caption{\textsc{Inferred graph metrics for each method}}
\begin{tabular}{|l|l|l|l|l|l|l|l|l|}
\hline
\textbf{Graph Type} & \textbf{\% Feasible Episodes} & \textbf{\#Edges} & \textbf{Avg out-deg.} & \textbf{Max out-deg.} & \textbf{Max in-deg.} & \textbf{Diameter} & \textbf{Avg shortest path} & \textbf{\#Scc} \\ \hline
\textbf{\texttt{Constrained-EM}}  & 98.90 & 100,073 & 2.38  & 448 & 133  & 28  & 5.95 & 17  \\ \hline
\textbf{\texttt{Saito et al.} \cite{b5}} & 10.06 & 20,542  & 0.49 & 21  & 18 & 88  & 8.78  & 0   \\ \hline
\textbf{\texttt{Star}}  & 100.00  & 162,570 & 3.87  & 4,524  & 173  & 18  & 6.03 & 23  \\ \hline
\textbf{\texttt{Chain}}  & 100.00  & 194,964 & 4.64  & 224 & 252  & 183 & 6.66  & 12  \\ \hline
\textbf{\texttt{Newman} \cite{b7}}   & 0.48 & 21,431  & 0.51 & 59  & 1,481 & 12  & 4.55  & 8  \\ \hline
\end{tabular}
\label{tab4}
\vspace{-5mm}
\end{table*}

\subsection{Comparison}

To generate the hidden friendship network $G$ inferred by \texttt{Constrained-EM}, we round up the edges $(i,j)$ with $Q_{ij} > 0.5$ to $1$, and the edges with $Q_{ij} <= 0.5$ to $0$. Each existing $(i,j)$ edge suggests that user $j$ follows user $i$. Then, we compare \texttt{Constrained-EM} with the following inference methods:

\begin{itemize}
  \item \textbf{\texttt{Star}}: a baseline inference method that creates an edge from the user who originally posted each tweet $s$ in the trace, to every other user who retweeted it.
  \item \textbf{\texttt{Chain}}: a baseline inference method which, for each episode $\Ec_{s}$ in the trace, creates a single long path between the user nodes in $\Ec_{s}$ according to the timestamps of their actions: from the user who originally posted $s$, to the user who retweeted $s$ first, to the next user who retweeted $s$, and so on.
  \item \textbf{\texttt{Saito et al.} \cite{b5}}: an EM-based algorithm that considers the friendship graph as a pre-given and infers the influence probabilities $k_{ij}$. For evaluation, we create a graph by drawing an edge $(i,j)$ whenever $k_{ij} > 0.5$. 
  \item \textbf{\texttt{Newman} \cite{b7}}: the EM-based algorithm by Newman, presented in the introduction. It is not designed to consider hidden paths between user tweets and retweets and therefore infers networks that are not feasible. However, it would be useful to observe the differences with our method. For evaluation, we create a graph by drawing an edge $(i,j)$ whenever the friendship probability $Q_{ij}$ for a user pair $(i,j)$ is  greater than  $0.5$.
\end{itemize}

To evaluate each method, since the ground truth is not available, we count how many episodes in the trace are feasible, given the graph inferred by each method. Moreover, we examine to what extent the properties of each graph resemble those of a real network. Therefore, we first look into the propagation graph inferred by each method, given a random episode from the trace $\Ec_{s}=\{1,2,...,9\}$ (users are anonymized by integers) and demonstrate it in Fig. \ref{fig:example}. Then, we compare each method on different graph statistics (Table \ref{tab4}). On top of that, we compare the in and out-degree complementary cumulative distribution functions (CCDFs) of each graph (Fig. \ref{fig:cdd}). From these, we can make the following observations for each method: 
\subsubsection{\texttt{Star}} Fig. \ref{fig:example} shows that the graph inferred by \texttt{Star} explains the whole episode $\Ec_{s}$ by connecting the author directly to each user that retweeted it. This is repeated for all episodes in the trace, achieving $100\%$ feasibility, as shown in Table \ref{tab4}. However, the way nodes are connected is heuristic and untrustworthy. This is also reflected in the unrealistically high maximum out-degree of its graph ($=4,524$), compared to the other methods.

\subsubsection{\texttt{Chain}} Fig. \ref{fig:example} shows that the graph inferred by \texttt{Chain} explains the whole episode $\Ec_{s}$ with a $100\%$ feasibility rate for all episodes. Again, \texttt{Chain} may return a feasible graph, but its high diameter $(=183)$ prevents us from choosing it as a real-world scenario. 
\subsubsection{\texttt{Saito et al.} \cite{b5}} In Fig. \ref{fig:example} we observe that for episode $\Ec_{s}$, the inferred graph explains only the retweet by user $2$ (through the $(1,2)$ edge). For the whole trace, it explains only $10\%$ of it (Table \ref{tab4}). Moreover, we observe that the graph has no strongly connected components (\#Scc in Table \ref{tab4}) which could also explain the small number of feasible episodes.
\subsubsection{\texttt{Newman} \cite{b7}} The vanilla method by Newman that does not use any constraints, cannot explain any interaction in episode $\Ec_{s}$, as expected. For the whole trace, it explains less than $0.50\%$ of the episodes.
\subsubsection{\texttt{Constrained-EM}} 
\begin{figure}
  \subfloat{{\includegraphics[width=4.5cm]{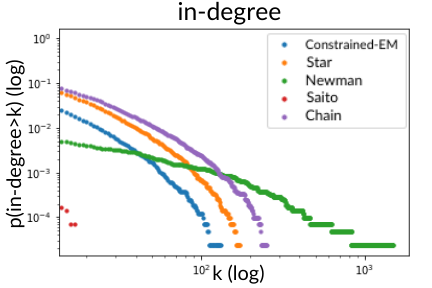} }}%
  \subfloat{{\includegraphics[width=4.5cm]{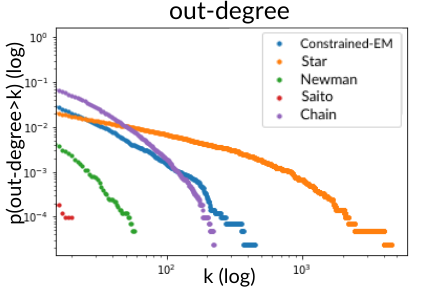} }}%
  \caption{CCDF for all methods (on a log-log scale).}%
  \label{fig:cdd}%
\end{figure}

Fig. \ref{fig:example} shows that our method can explain the whole $\Ec_{s}$, while for all episodes it achieves an almost $99\%$ feasibility for the given convergence rule (Table \ref{tab4}). The remaining $1\%$ that is left unexplained is attributed to the (rare) case when in an $\Ec_{s}$ there exist more than two users that have the same $M_{ij}$ values and thus, are not distinguished with sufficient certainty ($\sigma_{ij}=0.5$). We underline that our method is feasible with the least number of edges $(=100,073)$ compared to \texttt{Star} and \texttt{Chain}. Moreover, in Fig. \ref{fig:cdd}, especially in out-degree CCDF, \texttt{Constrained-EM} presents a close to \textit{scale-free} behavior, with both a heavy tail and an almost linear distribution line. On top of that, the average shortest path of our graph is close to 6. Given the well-known notion of \textit{six degrees of separation}, or equivalently, the idea that in a \textit{small-world} graph, any two pairs of nodes are separated by less than six nodes \cite{b17}, we conclude that our graph has properties close to these of a scale-free, small-world network. Hence, we consider it to be a trustworthy framework for feasible network inference. 

\section{Conclusion and Future work}

As demonstrated above, given a log of tweets and retweets, our method \texttt{Constrained-EM} successfully infers a feasible friendship graph that explains each tweet's propagation from user to user, while being economical in the number of
drawn edges. On top of that, we showed that our graph has properties that are close to these of a scale-free, small-world network. Therefore, \texttt{Constrained-EM} generates feasible graphs that are more reasonable than simple heuristics like \texttt{Star} and \texttt{Chain}. It is worth noting that our method could be applied on other domains where feasibility constraints can be imposed, such as epidemics, biology, etc. As future work, we plan to investigate ways in which \texttt{Constrained-EM} can be improved in terms of convergence speed.

\vspace{12pt}
\end{document}